\documentclass[a4paper,11pt]{article}
\usepackage{pos}
\usepackage{lipsum} 
\usepackage{setspace}

\title{Progress towards an array-wide diffuse UHE neutrino search with the Askaryan Radio Array}
\ShortTitle{Progress towards an array-wide diffuse UHE neutrino search with ARA}

\author*[a,b,c]{Marco Stein Muzio}
\onbehalf{for the ARA Collaboration \\{\normalsize \normalfont(a complete list of authors can be found at the end of the proceedings)}\\}

\affiliation[a]{Center for Multi-Messenger Astrophysics, Institute for Gravitation and the Cosmos,\\ Pennsylvania State University, University Park, PA 16802, USA}

\affiliation[b]{Department of Physics, Pennsylvania State University, \\University Park, PA 16802, USA}

\affiliation[c]{Department of Astronomy and Astrophysics, Pennsylvania State University, \\University Park, PA 16802, USA}

\emailAdd{msm6428@psu.edu}

\abstract{

The Askaryan Radio Array (ARA) is an in-ice ultrahigh energy (UHE) neutrino experiment at the South Pole. ARA aims to detect the radio emissions from neutrino-induced particle showers using in-ice clusters of antennas buried ${\sim}200$~m deep on a roughly cubical lattice with side length of ${\sim}10$~m. ARA has five such independent stations which have collectively accumulated ${\sim}30$~station-years of livetime through 2023. The fifth station of ARA has an additional sub-detector, known as the phased array, which pioneered an interferometric trigger constructed by beamforming the signals of $7$~tightly packed, vertically-polarized antennas. This scheme has been demonstrated to significantly improve the trigger efficiency for low SNR signals. In this talk, we will present the current state of the first array-wide diffuse neutrino search using $24$~station-years of data (through 2021). We anticipate that this analysis will result in the first UHE neutrino observation or world-leading limits from a radio neutrino detector below $100$~EeV. Additionally, this analysis will demonstrate the feasibility for multi-station in-ice radio arrays to successfully conduct an array-wide neutrino search --- paving the way for future, large detector arrays such as RNO-G and IceCube-Gen2 Radio.

}

\FullConference{10th International Workshop on Acoustic and Radio EeV Neutrino Detection Activities (ARENA2024)\\
11-14 June 2024\\
The Kavli Institute for Cosmological Physics, Chicago, IL, USA\\}

\begin{document}
\maketitle

\section{Introduction}\label{sec:intro}

\par
Ultrahigh energy cosmic rays (UHECRs) have been robustly detected with energies in excess of $100$~EeV~$=10^{20}$~eV, yet both their sources and the astrophysical acceleration mechanisms which can realize such enormous energies remain an open question. However, Nature provides a unique lens through which to study UHECRs and their sources: cosmic neutrinos. UHECRs are expected to produce $\gtrsim10$~PeV neutrinos through interactions in their source environment, as well as, through interactions with the cosmic microwave background (CMB) \& extragalactic background light (EBL) during extragalactic propagation. In particular, neutrinos produced extragalactically via interactions between UHECRs $\gtrsim 50$~EeV and the CMB are expected to have energies $\gtrsim1$~EeV. These neutrinos are produced via $\Delta$ resonance production which significantly degrades the UHECR energy, leading to a suppression of the UHECR spectrum. This effect is often called the Greisen-Zatsepin-Kuzmin (GZK) cutoff~\cite{Greisen:1966jv,Zatsepin:1966jv}. Neutrinos themselves are ideal cosmic messengers, since they probe cosmological scales and are not deflected by intervening magnetic fields. This makes them an indispensable tool for identifying UHECR sources and probing  UHECR source properties beyond the GZK horizon. 

\par
Neutrinos interacting in a dense dielectric medium will initiate a particle cascade which will produce a sub-nanosecond pulse of coherent radio emission along the Cherenkov cone, often called Askaryan emission~\cite{Askaryan:1961pfb,ANITA:2006nif}. Radio waves have $\mathcal{O}(\mathrm{km})$ attenuation lengths in glacial ice~\cite{Barwick:2005zz}, meaning that in-ice radio detectors are able to efficiently monitor enormous $\mathcal{O}(\mathrm{km}^3)$ volumes of ice for such interactions. This is fortuitous since teraton-scale detectors are required to observed neutrinos at these energies --- only $\mathcal{O}(10^{-2})$ neutrino interactions are expected in a cubic kilometer of ice per year per energy decade, due to the small neutrino flux expected beyond ${\sim}10$~PeV~\cite{Ackermann:2022rqc} and small neutrino-nucleon cross-section~\cite{Connolly:2011vc}.

\par
A sparse array of independent radio detector stations makes it possible to efficiently build a roughly $100$~km$^3$ water-equivalent detector. A number of current and upcoming detectors leverage this concept including the Askaryan Radio Array (ARA), ARIANNA~\cite{Barwick:2016mxm}, RNO-G~\cite{RNO-G:2020rmc}, and IceCube-Gen2 Radio~\cite{IceCubeGen2_TDR}. Among these ARA has the longest livetime and, in this proceeding, we report on the progress towards conducting the first array-wide search for UHE neutrinos.

\section{ARA Detector Overview}\label{sec:ARA}

\begin{figure}[h!]
  \centering
  \includegraphics[width=0.49\textwidth]{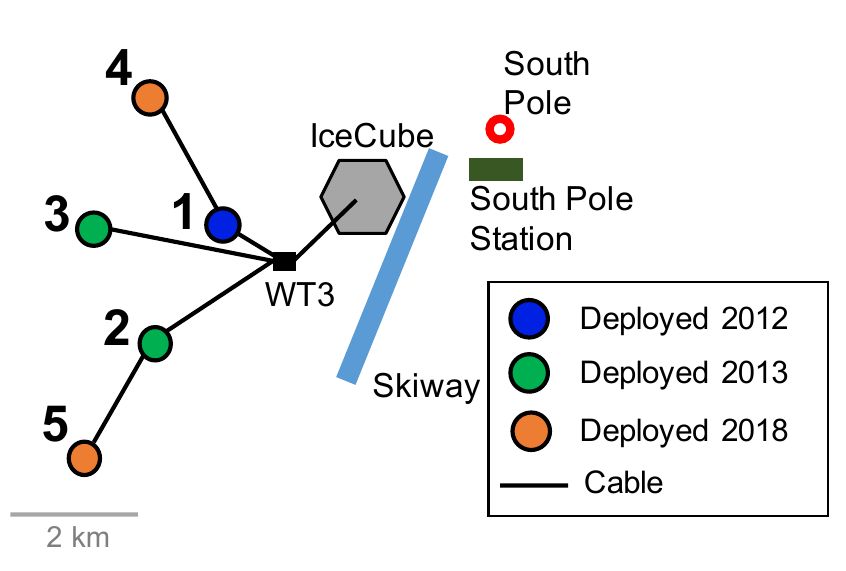}
  \hfill
  \includegraphics[width=0.49\textwidth]{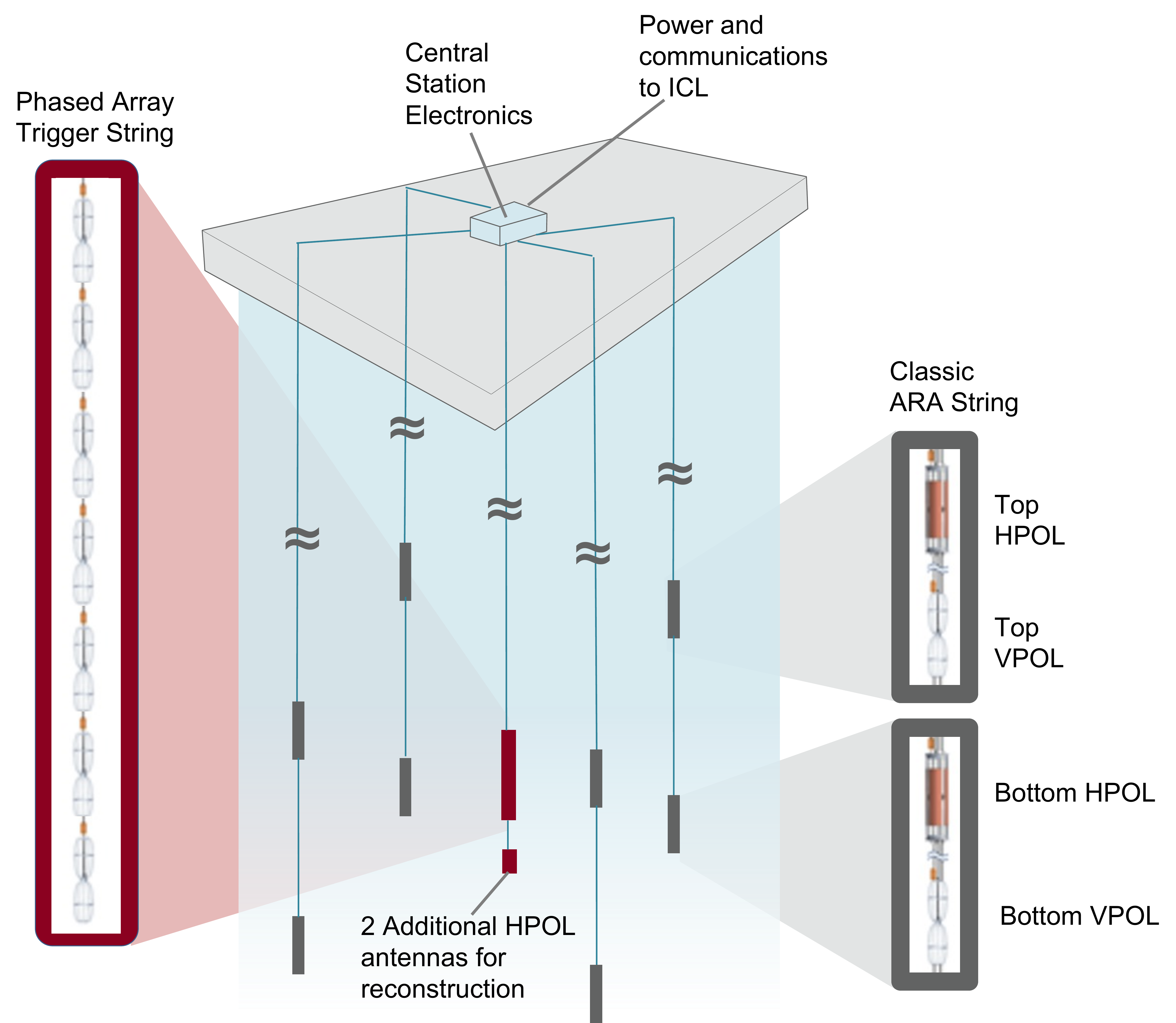}
  \caption{Left: Layout of the ARA station array, relative to IceCube and the South Pole. Right: Layout of the two subdetectors of A5. Omitting the central PA string (red), other stations of ARA (A1-4) have the same layout.}
  \label{fig:ara_schematic}
\end{figure}

The ARA detector is located at the South Pole and consists of five independent stations (A1-5, collectively ARA5) on a hexagonal grid with a $2$~km spacing (see Fig.~\ref{fig:ara_schematic}). Each of these stations has four strings of antennas, each deployed with a maximum depth of roughly $200$~m below the surface of the ice. Each string has four antennas, two primarily sensitive to vertically-polarized signals (VPols) and two to horizontally-polarized signals (HPols). These antennas are sensitive to signals from $150-850$~MHz~\cite{ARA:2019wcf}. This particular detector layout was chosen to optimize the effective area to Askaryan signals from neutrinos $E_\nu > 10^{18}$~eV. ARA stations trigger when three antennas of the same polarization record an integrated power over a $25$~ns window five times above the ambient noise level within an approximately $170$~ns coincidence window. This results in a ${\sim}6$~Hz trigger rate, including a $1$~Hz calibration pulser, and an additional $1$~Hz forced trigger used to monitor the ambient noise environment.

\par
In addition to these four ``measurement'' strings, each ARA station has one or two additional strings for calibration. Each of these ``calibration'' strings are located ${\sim}40$~m from station center and have two transmitter antennas, one VPol and one HPol, deployed at comparable depths to the measurement strings. These transmitter antennas can emit both broadband and continuous wave signals for in-situ position and timing calibration. 

\par
The fifth station of ARA, A5, has an additional subdetector called the phased array (PA). The PA consists of a single string of seven closely packed VPol antennas, as well as two HPol antennas, deployed at station-center. The PA coherently sum the signals in each VPol channel along $15$~pre-defined directions, called beams, before the trigger condition is checked. This makes the PA more sensitive to low SNR signals since signals will coherently sum whereas noise will add incoherently, boosting the SNR at the trigger. In additional to this improvement in trigger efficiency, the PA has also been shown to have an improved analysis efficiency in neutrino-background signal discrimination~\cite{ARA:2022rwq}. The PA has a separate DAQ from the traditional ARA detector on A5, which triggers when a beam has excess power in a $10$~ns window over a threshold adjusted to maintain an $11$~Hz global trigger rate across all beams. Notably, the traditional ARA DAQ on A5 was lost at the end of 2019 due to a USB port failure. For this reason, from 2020 onwards $6$~additional VPols from the traditional station were connected to the PA DAQ (limited by the number of open channels on this DAQ). This merged system is referred to as A5/PA to distinguish it from when both systems were totally independent. The ARA station layout, for the particular case of A5, is depicted in Fig.~\ref{fig:ara_schematic}, with calibration string omitted for clarity.

\section{An Array-Wide Neutrino Search}\label{sec:arrayWide}

\par
The five stations of ARA were deployed between 2012 and 2018. Through 2021 approximately $24$~station-years of livetime and $310$~TB of data have been accumulated. An array-wide neutrino search across this entire livetime is required to realize the full potential of this unprecedented data set and ARA's full sensitivity. Previous searches in the ARA data have analyzed approximately $6.6$~station-years of livetime --- less than $30\%$ of the total livetime available~\cite{ARA:2019wcf,ARA:2022rwq}. Moreover, this pioneering search will demonstrate the feasibility for multi-station in-ice radio arrays to successfully conduct an array-wide neutrino search. The ability to successfully carry out such a search is critical for future large-scale arrays, such as RNO-G ($35$~stations) and IceCube-Gen2 Radio ($361$~stations), to reach their design sensitivity.

\par
In order to maximize the sensitivity of such a search, the final selection criteria to minimize backgrounds and maximize potential signals must be optimized across the full array, rather than station-by-station. However, each station is independent with its own set of backgrounds which must be characterized and mitigated against. Therefore, the analysis is split into two parts: station-level analyses and a final global optimization. Station-level analyses focus on the data and background characterization in a single station, as well as, development of cuts to remove non-thermal backgrounds based on a $10\%$ burn sample. Given their scale, the station-level analyses are being carried out by a core team of seven analyzers via a highly-coordinated effort  across nine institutions within the ARA Collaboration. 

\begin{figure}[h!]
  \centering
  \includegraphics[width=0.625\textwidth]{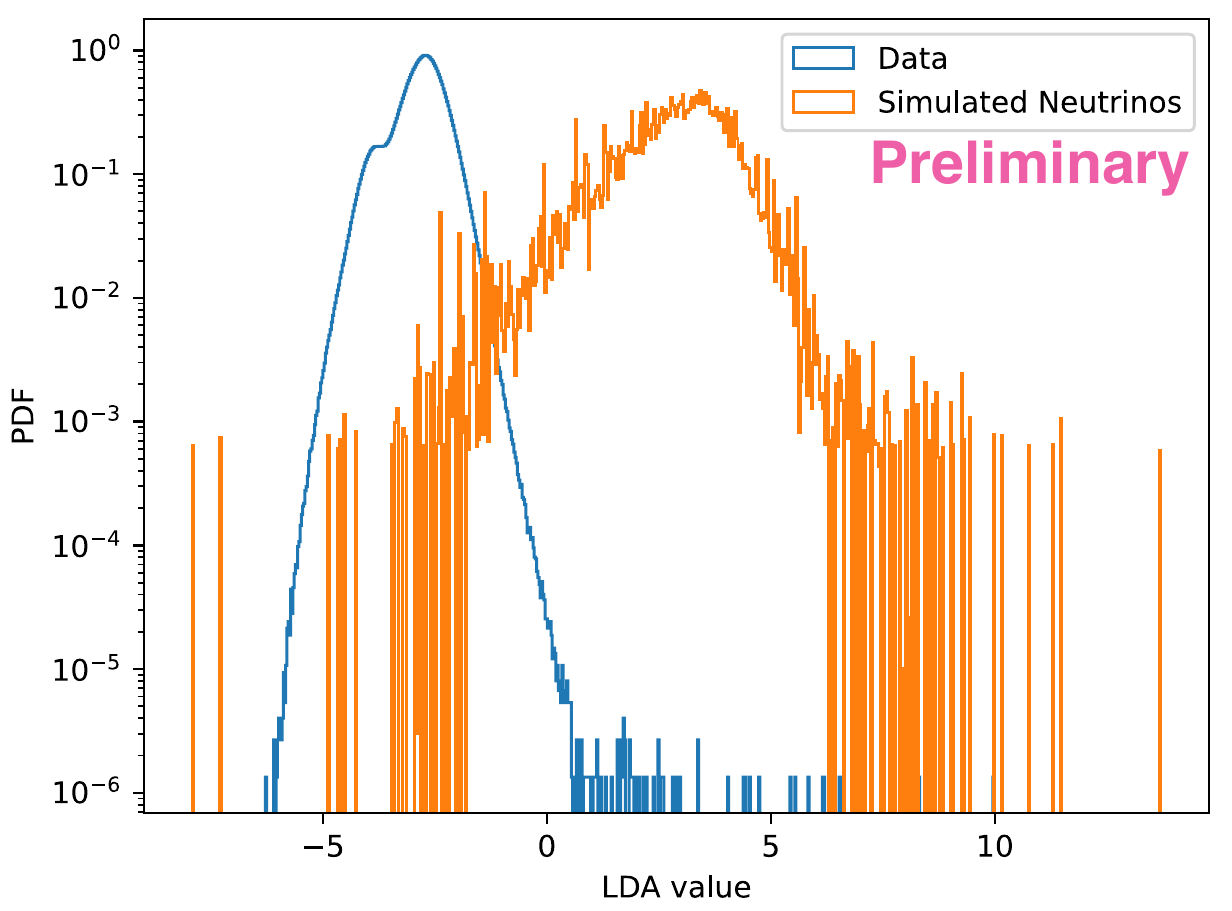}
  \caption{Example of separation between background events and simulated neutrinos in terms of LDA value (referred to as $t_s$ in the text) for the A5/PA burn sample. Note that the data distribution is thermal noise dominated but non-thermal events have not been fully removed and make up the high-LDA value tail.}
  \label{fig:lda_example}
\end{figure}

\par
The global optimization is the final step of the analysis. Station-level analyses will produce a set of cuts to remove non-thermal backgrounds from their burn sample. In addition, they will characterize the remaining data, as well as, a set of simulated neutrino events in terms of a set of selection variables. Each of these selection variables (e.g. signal-to-noise ratio, maximum correlation between channel signals, impulsivity of the signal, etc.) have different signal-background discrimination power. In order to maximize separation of simulated neutrino signals from thermal background signals, a linear discriminant analysis (LDA) is used to combine these selection variables into a single, final selection variable $t_s$ (see Fig.~\ref{fig:lda_example} for an example). Station-level analyses pass their signal efficiency $\epsilon_s$ (i.e.~the fraction of triggered neutrinos in the signal region) and background leakage rate $b_s$ (i.e.~the number of background events in the signal region) as a function of $t_s$, as well as, their livetime $T_s$ and effective area $A_{\mathrm{eff},s}$ to the global optimization. The global optimization then performs a final tuning of $t_s$ to optimize the threshold for the signal region for $5\sigma$ discovery potential across the array. This is very similar to (the simpler) optimization rule for the strongest upper limit on the neutrino flux, 

\begin{align}
    \phi_\mathrm{UL} = \frac{\mathrm{FC}(\sum_s b_s(t_s))}{\sum_s A_{\mathrm{eff}, s} T_s \epsilon_s(t_s)}~,
\end{align}

\noindent
where $\mathrm{FC}$ is the Feldman-Cousins $90\%$ confidence level upper limit on the true number of neutrino events assuming no events are observed given a background leakage of $\sum_s b_s$. To estimate $b_s$ and $\epsilon_s$ we assume the true neutrino flux is given by the $90\%$ confidence level upper limit on the UHE diffuse neutrino flux from IceCube~\cite{IceCube:2018fhm}. This flux was chosen since it represents the maximum flux allowed by current data, agnostic to any particular theoretical model. Performing the final optimization over the entire array enables a more sensitive search than possible optimizing station-by-station.

\par
\begin{figure}[h!]
  \centering
  \includegraphics[width=0.625\textwidth]{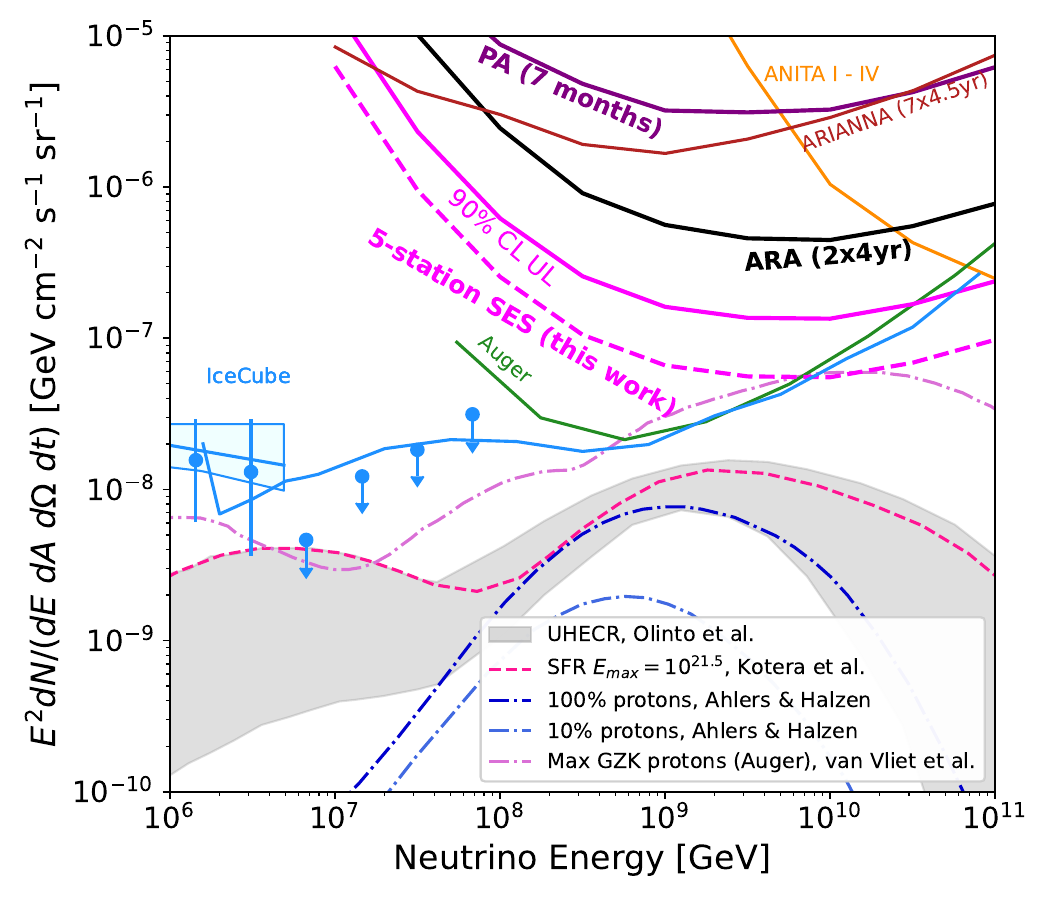}
  \caption{Projected sensitivity at analysis level for the array-wide search in terms of the projected $90\%$~confidence level upper limit (solid) and single-event sensitivity (dashed). Data, limits, and theoretical predictions are also shown for reference.}
  \label{fig:projected_limit}
\end{figure}

The result of this two part analysis will either be the first detection of UHE neutrino candidates, or ARA's strongest limit on the diffuse UHE neutrino flux. Up to a dozen events are expected in the analyzed data set at trigger level, under a number of UHE neutrino flux models: $12.5$~events are expected under the IceCube flux limit model~\cite{IceCube:2018fhm}; $9.7$~events are expected under the Auger-tuned flux model from~\cite{vanVliet:2019nse,Anker:2020lre}; and, $2.1$~events are expected under the flux model of~\cite{Kotera:2010yn}. Under these models it is plausible that this analysis could result in a UHE neutrino candidate. However, in the case that no neutrino candidates are found the projected upper-limit on the diffuse UHE neutrino flux will be the strongest set by any in-ice radio detector up to $100$~EeV (see Fig.~\ref{fig:projected_limit}).

\section{Improvements to Analysis \& Simulation}\label{sec:improvements}

\par
In support of this array-wide neutrino search, a number of improvements have been made to both the characterization of the detector in simulation, as well as, improvements to background separation in analysis. These improvements have helped to increase detector simulation fidelity, made more accurate estimates of the effective area of each station, and have improved the overall sensitivity of station-level analyses. 

\par
Three major improvements have been made to ARA's detector characterization in simulation. First, data-driven noise models have been developed and incorporated into simulation. Data-driven noise models were generated by analyzing forced trigger events in each station. For each event, spectral amplitude in each frequency bin was recorded for each channel. From these a distribution of spectral amplitudes for every frequency bin was produced. These were fit to a Rayleigh distribution and their widths (or spectral coefficients) were recorded. These spectral coefficients (for each frequency and channel on a station) were incorporated into the ARA simulation package, \verb|AraSim|~\cite{ARA:2014fyf}. To produce a noise waveform in simulation, the spectral amplitude is drawn from Rayleigh distributions of the recorded width and a random phase is drawn for each frequency. 

\par
\begin{figure}[h!]
  \centering
  \includegraphics[width=0.49\textwidth]{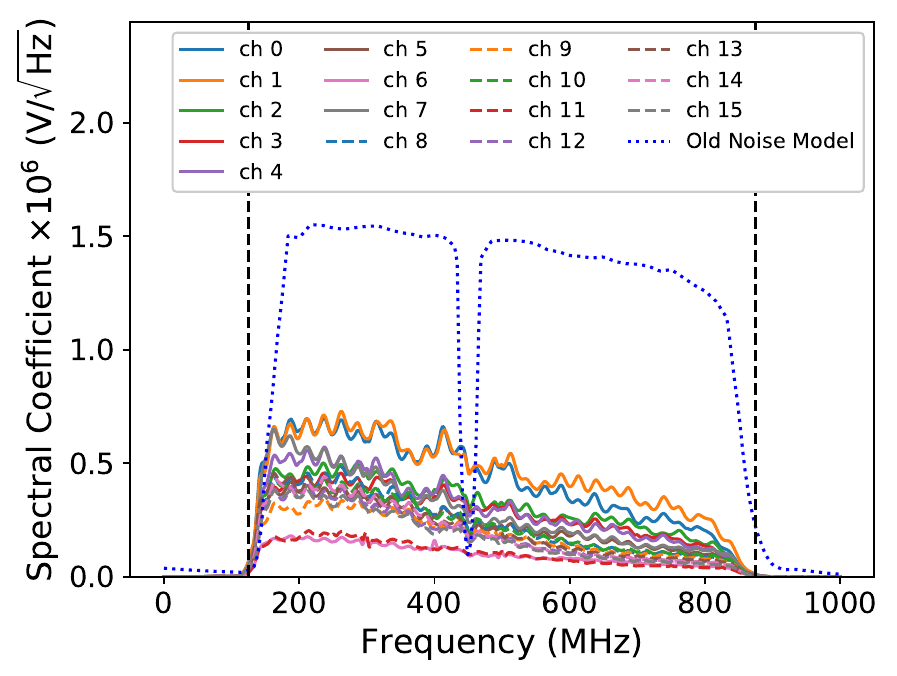}
  \hfill
  \includegraphics[width=0.49\textwidth]{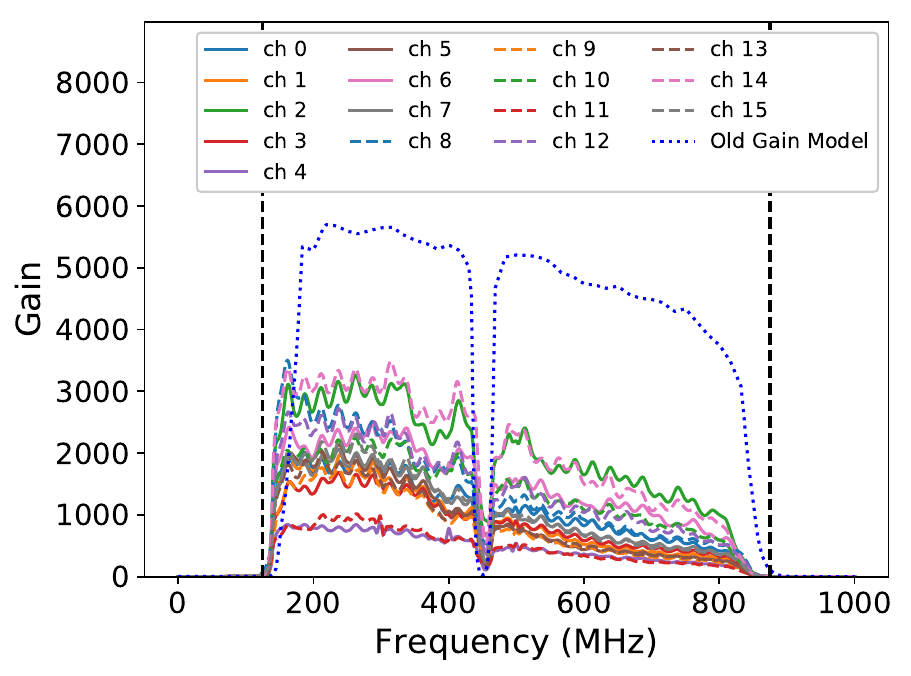}
  \caption{Per-channel data-driven noise (left) and signal chain gain (right) models for the first livetime configuration of A3. The previously used models are shown in blue dotted lines. The ARA frequency band is denoted by dashed vertical lines.}
  \label{fig:noiseGainModels}
\end{figure}

Second, data-driven signal chain gain (amplitude) models have also been developed and incorporated into simulation. These gain models were constructed by taking the ratio of the spectral widths of software triggers (i.e. the data-driven noise models) to the theoretical expectation for thermal noise given the ice temperature and antenna \& amplifier properties. Both data-driven noise and signal chain gain models were constructed for each stable livetime period for a station. A comparison of these data-driven models to previous models can be found in Fig.~\ref{fig:noiseGainModels}.

\par
\begin{figure}[h!]
  \centering
  \includegraphics[width=\textwidth]{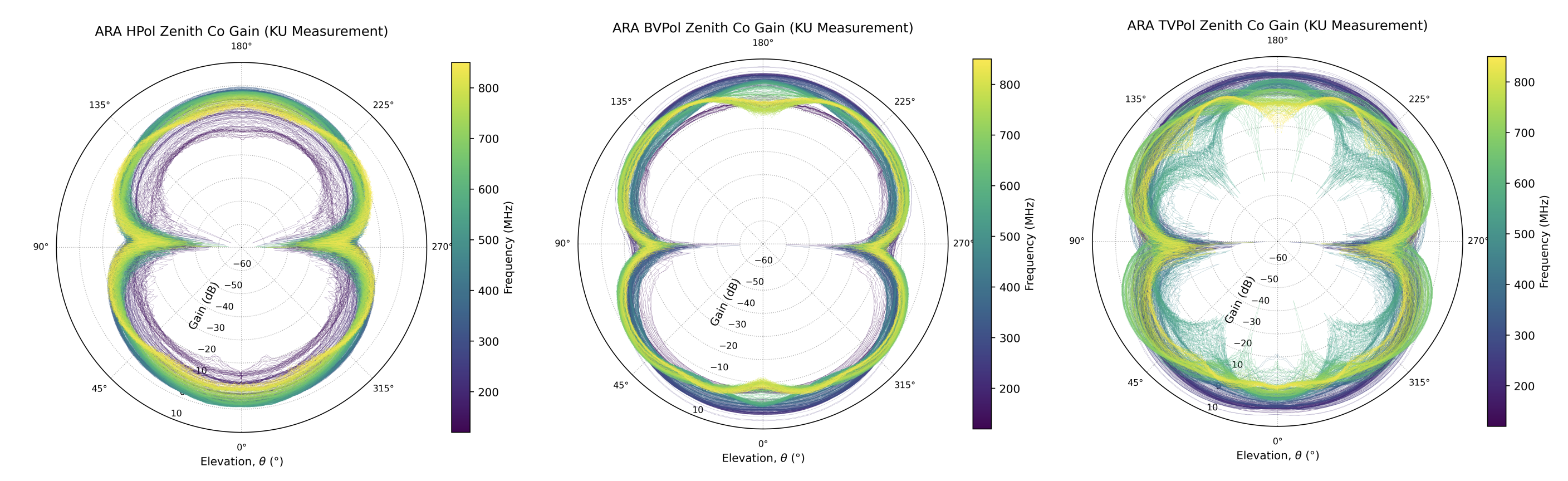}
  \caption{Measured antenna gain pattern for HPols (left), bottom (center) \& top (right) VPols.}
  \label{fig:antennaGains}
\end{figure}

Finally, new anechoic chamber measurements have been made at University of Kansas. In particular, these measurements were taken for the three distinct antenna types used on ARA stations: HPols (quadslot antennas) and top \& bottom VPols (dipole antennas). The properties of top and bottom VPols are distinct due to their differing number of through-cables. The frequency-dependent gain pattern measured for each of these antenna types can be seen in Fig.~\ref{fig:antennaGains}.

\par
\begin{figure}[h!]
  \centering
  \includegraphics[width=0.625\textwidth]{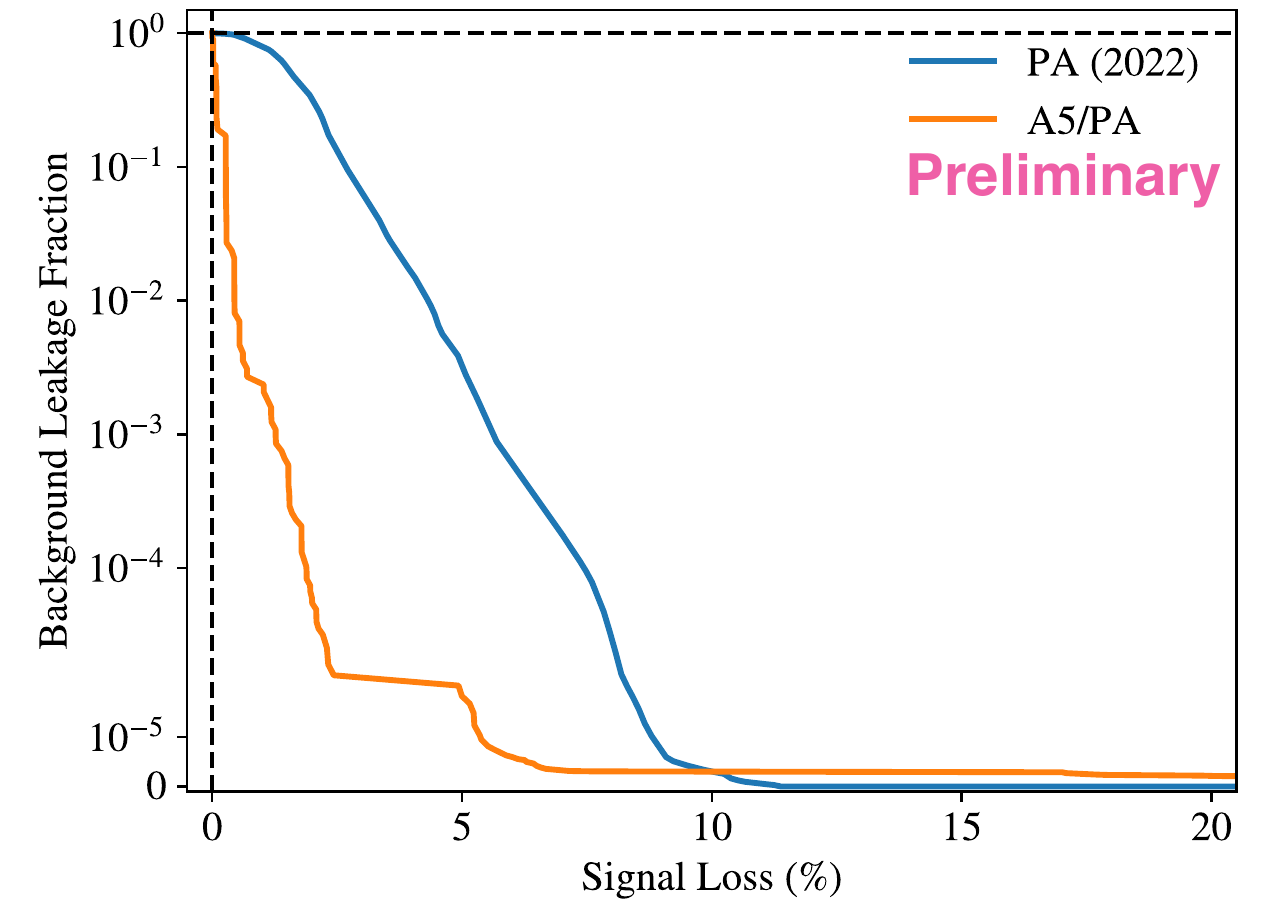}
  \caption{Comparison of the background leakage fraction for a given signal loss rate for the current A5/PA analysis compared to the previous PA-only analysis~\cite{ARA:2022rwq}. Note that non-thermal events have not been fully cleaned from the A5/PA burn sample, leading to a long, non-zero tail in this plot.}
  \label{fig:backgroundSeparation}
\end{figure}

Improvements to both thermal and non-thermal background discrimination from signal have also been made. Improvements to event characterization, in particular implementation of waveform de-dispersion and coherently summing waveforms across all channels, have improved analysis efficiency by improving discrimination of neutrino signals from thermal backgrounds. Non-thermal background suppression, in particular in the PA data set, has also been significantly improved through implementation of spatio-temporal clustering cuts. Figure~\ref{fig:backgroundSeparation} demonstrates these improvements for the combined A5/PA detector compared to the previous PA analysis~\cite{ARA:2022rwq}.

\section{Conclusion}\label{sec:conclusion}

\par
The ARA detector has been taking data since 2012. Through 2021 ARA has accumulated roughly $24$~station-years of livetime across its five station array. This represents the largest exposure of any in-ice radio detector. To fully leverage this data the ARA Collaboration has embarked on a pioneering, multi-institution, array-wide search for diffuse UHE neutrinos in this full data set. This analysis will be the first to demonstrate the feasibility of an array-wide search, and will be a key proof-of-concept for upcoming large radio arrays, such as RNO-G and IceCube-Gen2 Radio. 

\par
The analysis is split into two parts, station-level analyses and final global optimizations. Station-level analyses will remove and characterize all backgrounds in a $10\%$~burn sample. These will produce a single, linear discriminant analysis value to characterize data and simulated events with maximum discrimination power. The final global optimization will define the signal threshold in terms of this final variable to optimize thermal background removal for $5\sigma$~discovery potential across the array. In addition to carrying out this analysis, a number of significant improvements have been made to the ARA detector characterization (including new data-driven noise \& signal chain gain models, as well as, new antenna measurements) and analysis efficiency.

\par
This analysis has significant potential to discover the first UHE neutrino candidate. If no such candidate is found, this analysis will produce the strongest upper-limit on the UHE neutrino flux below $100$~EeV of any in-ice radio experiment --- having also paved the way for future large-array experiments to fully realize their design sensitivity.

\begingroup
\setstretch{0.5}
\setlength{\bibsep}{1.0pt}
\bibliographystyle{JHEP}
\bibliography{ARENA24_Muzio_FiveStation}
\endgroup

%

\clearpage

\section*{Full Author List: ARA Collaboration (July 22, 2024)}

\noindent
S.~Ali\textsuperscript{1},
P.~Allison\textsuperscript{2},
S.~Archambault\textsuperscript{3},
J.J.~Beatty\textsuperscript{2},
D.Z.~Besson\textsuperscript{1},
A.~Bishop\textsuperscript{4},
P.~Chen\textsuperscript{5},
Y.C.~Chen\textsuperscript{5},
Y.-C.~Chen\textsuperscript{5},
B.A.~Clark\textsuperscript{6},
A.~Connolly\textsuperscript{2},
K.~Couberly\textsuperscript{1},
L.~Cremonesi\textsuperscript{7},
A.~Cummings\textsuperscript{8,9,10},
P.~Dasgupta\textsuperscript{2},
R.~Debolt\textsuperscript{2},
S.~de~Kockere\textsuperscript{11},
K.D.~de~Vries\textsuperscript{11},
C.~Deaconu\textsuperscript{12},
M.~A.~DuVernois\textsuperscript{4},
J.~Flaherty\textsuperscript{2},
E.~Friedman\textsuperscript{6},
R.~Gaior\textsuperscript{3},
P.~Giri\textsuperscript{13},
J.~Hanson\textsuperscript{14},
N.~Harty\textsuperscript{15},
K.D.~Hoffman\textsuperscript{6},
M.-H.~Huang\textsuperscript{5,16},
K.~Hughes\textsuperscript{2},
A.~Ishihara\textsuperscript{3},
A.~Karle\textsuperscript{4},
J.L.~Kelley\textsuperscript{4},
K.-C.~Kim\textsuperscript{6},
M.-C.~Kim\textsuperscript{3},
I.~Kravchenko\textsuperscript{13},
R.~Krebs\textsuperscript{8,9},
C.Y.~Kuo\textsuperscript{5},
K.~Kurusu\textsuperscript{3},
U.A.~Latif\textsuperscript{11},
C.H.~Liu\textsuperscript{13},
T.C.~Liu\textsuperscript{5,17},
W.~Luszczak\textsuperscript{2},
K.~Mase\textsuperscript{3},
M.S.~Muzio\textsuperscript{8,9,10},
J.~Nam\textsuperscript{5},
R.J.~Nichol\textsuperscript{7},
A.~Novikov\textsuperscript{15},
A.~Nozdrina\textsuperscript{1},
E.~Oberla\textsuperscript{12},
Y.~Pan\textsuperscript{15},
C.~Pfendner\textsuperscript{18},
N.~Punsuebsay\textsuperscript{15},
J.~Roth\textsuperscript{15},
A.~Salcedo-Gomez\textsuperscript{2},
D.~Seckel\textsuperscript{15},
M.F.H.~Seikh\textsuperscript{1},
Y.-S.~Shiao\textsuperscript{5,19},
S.C.~Su\textsuperscript{5},
S.~Toscano\textsuperscript{20},
J.~Torres\textsuperscript{2},
J.~Touart\textsuperscript{6},
N.~van~Eijndhoven\textsuperscript{11},
G.S.~Varner\textsuperscript{21,$\dagger$},
A.~Vieregg\textsuperscript{12},
M.-Z.~Wang\textsuperscript{5},
S.-H.~Wang\textsuperscript{5},
S.A.~Wissel\textsuperscript{8,9,10},
C.~Xie\textsuperscript{7},
S.~Yoshida\textsuperscript{3},
R.~Young\textsuperscript{1}
\\
\\
\textsuperscript{1} Dept. of Physics and Astronomy, University of Kansas, Lawrence, KS 66045\\
\textsuperscript{2} Dept. of Physics, Center for Cosmology and AstroParticle Physics, The Ohio State University, Columbus, OH 43210\\
\textsuperscript{3} Dept. of Physics, Chiba University, Chiba, Japan\\
\textsuperscript{4} Dept. of Physics, University of Wisconsin-Madison, Madison,  WI 53706\\
\textsuperscript{5} Dept. of Physics, Grad. Inst. of Astrophys., Leung Center for Cosmology and Particle Astrophysics, National Taiwan University, Taipei, Taiwan\\
\textsuperscript{6} Dept. of Physics, University of Maryland, College Park, MD 20742\\
\textsuperscript{7} Dept. of Physics and Astronomy, University College London, London, United Kingdom\\
\textsuperscript{8} Center for Multi-Messenger Astrophysics, Institute for Gravitation and the Cosmos, Pennsylvania State University, University Park, PA 16802\\
\textsuperscript{9} Dept. of Physics, Pennsylvania State University, University Park, PA 16802\\
\textsuperscript{10} Dept. of Astronomy and Astrophysics, Pennsylvania State University, University Park, PA 16802\\
\textsuperscript{11} Vrije Universiteit Brussel, Brussels, Belgium\\
\textsuperscript{12} Dept. of Physics, Enrico Fermi Institue, Kavli Institute for Cosmological Physics, University of Chicago, Chicago, IL 60637\\
\textsuperscript{13} Dept. of Physics and Astronomy, University of Nebraska, Lincoln, Nebraska 68588\\
\textsuperscript{14} Dept. Physics and Astronomy, Whittier College, Whittier, CA 90602\\
\textsuperscript{15} Dept. of Physics, University of Delaware, Newark, DE 19716\\
\textsuperscript{16} Dept. of Energy Engineering, National United University, Miaoli, Taiwan\\
\textsuperscript{17} Dept. of Applied Physics, National Pingtung University, Pingtung City, Pingtung County 900393, Taiwan\\
\textsuperscript{18} Dept. of Physics and Astronomy, Denison University, Granville, Ohio 43023\\
\textsuperscript{19} National Nano Device Laboratories, Hsinchu 300, Taiwan\\
\textsuperscript{20} Universite Libre de Bruxelles, Science Faculty CP230, B-1050 Brussels, Belgium\\
\textsuperscript{21} Dept. of Physics and Astronomy, University of Hawaii, Manoa, HI 96822\\
\textsuperscript{$\dagger$} Deceased\\

\section*{Acknowledgements}

\noindent
The ARA Collaboration is grateful to support from the National Science Foundation through Award 2013134.
The ARA Collaboration
designed, constructed, and now operates the ARA detectors. We would like to thank IceCube, and specifically the winterovers for the support in operating the
detector. Data processing and calibration, Monte Carlo
simulations of the detector and of theoretical models
and data analyses were performed by a large number
of collaboration members, who also discussed and approved the scientific results presented here. We are
thankful to Antarctic Support Contractor staff, a Leidos unit 
for field support and enabling our work on the harshest continent. We thank the National Science Foundation (NSF) Office of Polar Programs and
Physics Division for funding support. We further thank
the Taiwan National Science Councils Vanguard Program NSC 92-2628-M-002-09 and the Belgian F.R.S.-
FNRS Grant 4.4508.01 and FWO. 
K. Hughes thanks the NSF for
support through the Graduate Research Fellowship Program Award DGE-1746045. A. Connolly thanks the NSF for
Award 1806923 and 2209588, and also acknowledges the Ohio Supercomputer Center. S. A. Wissel thanks the NSF for support through CAREER Award 2033500.
A. Vieregg thanks the Sloan Foundation and the Research Corporation for Science Advancement, the Research Computing Center and the Kavli Institute for Cosmological Physics at the University of Chicago for the resources they provided. R. Nichol thanks the Leverhulme
Trust for their support. K.D. de Vries is supported by
European Research Council under the European Unions
Horizon research and innovation program (grant agreement 763 No 805486). D. Besson, I. Kravchenko, and D. Seckel thank the NSF for support through the IceCube EPSCoR Initiative (Award ID 2019597). M.S. Muzio thanks the NSF for support through the MPS-Ascend Postdoctoral Fellowship under Award 2138121. A. Bishop thanks the Belgian American Education Foundation for their Graduate Fellowship support.

\end{document}